\documentclass{osa-article}

\journal{osajournal}

\setprjcopyright

\articletype{Research Article}

\begin{document}

\title{Modal interferometric refractive index sensing in microstructured exposed core fibres}

\author{Ivan S. Maksymov,\authormark{1,*} Heike Ebendorff-Heidepriem,\authormark{2} and Andrew D. Greentree\authormark{3}}

\address{\authormark{1}Centre for Micro-Photonics, Swinburne University of Technology, Hawthorn, VIC 3122, Australia\\
\authormark{2}Australian Research Council Centre of Excellence for Nanoscale BioPhotonics, Institute for Photonics and Advanced Sensing, School of Physical Sciences, University of Adelaide, Adelaide, SA 5005, Australia\\
\authormark{3}Australian Research Council Centre of Excellence for Nanoscale BioPhotonics, School of Science, RMIT University, Melbourne, VIC 3001, Australia}

\email{\authormark{*}imaksymov@swin.edu.au} 



\begin{abstract}
Optical fibre-based sensors measuring refractive index shift in bodily fluids and tissues are versatile and accurate probes of physiological processes. Here, we suggest a refractive index sensor based on a microstructured exposed-core fibre (ECF). By considering a high refractive index coating of the exposed core, our modelling demonstrates the splitting of the guided mode into a surface sensing mode and a mode that is isolated from the surface. With the isolated mode acting as a reference arm, this two-mode one-fibre solution provides for robust interferometric sensing with a sensitivity of up to $60,000$\,rad/RIU-cm, which is suitable for sensing subtle physiological processes within hard-to-reach places inside living organisms, such as the spinal cord, ovarian tract and blood vessels.
\end{abstract}

\section{Introduction}
The analysis of physiological processes and early detection of certain diseases \cite{Wan11} requires sensors capable of probing subtle changes in temperature \cite{Cor18}, pH level \cite{Pur15, Che18}, and concentration of biological fluids and gases \cite{Pur15} in hard-to-reach places inside living organisms such as the brain \cite{Wel05}, spinal cord \cite{Lu14}, ovarian tract \cite{Wan15} and blood vessels \cite{review}. Such changes can often be sensed as small shifts in the optical refractive index of bodily fluids and tissues \cite{review}.  

\begin{figure}[t]
\centerline{
\includegraphics[width=8.5cm]{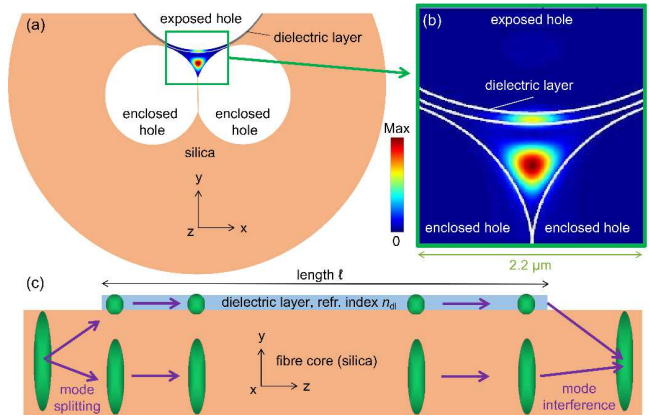}}
\caption{{\color{black}(a) Schematic of an ECF coated with a dielectric layer of thickness $h$, length $\ell$ [see (c)], and refractive index $n_{dl}$. (b) Optical intensity profile in the fibre with the $100$-nm-thick dielectric layer and $n_{dl}=2$. The fundamental guided mode confined in the Y-shaped core and a higher-order mode localised in the dielectric layer can be seen. (c) Longitudinal cross-section of the ECF and schematic of the mode behaviour. The mode of the bare ECF splits into the two guided modes due to the dielectric layer. The two modes co-propagate, re-couple and interfere near the output edge of the fibre.}}
\label{fig:fig1}
\end{figure}

Microstructured exposed-core optical fibres (ECFs) provide a broad range of optical properties demanded by biomedical refractive index sensors intended to operate inside living organisms \cite{War09, Kos12, Mes14, Bal17}. ECFs confine and guide light in a small volume of a dielectric material surrounded by longitudinal air holes (Fig.~\ref{fig:fig1}). One of the holes is open along the entire length of the fibre, which allows using it as a sample chamber where a portion of the guided light evanescently extends above the fibre and provides light-matter overlap and enhanced interaction required for sensing. 

Here, by theoretically considering a thin dielectric layer created on top of a standard ECF (Fig.~\ref{fig:fig1}), we show that the guided mode can be split into a surface sensing mode and a mode that is isolated from the surface. The isolated mode is immune to environmental changes, but the surface mode is highly sensitive to small refractive index shifts in the outer environment across the entire fibre length. We show that interference between the two modes can be used to create a fibre-optic sensor capable of detecting shifts in the optical refractive index with $60,000$\,rad/RIU-cm sensitivity. Such sensitivity is comparable with that of two-arm interferometers \cite{Hai09, Qin16}, two-mode one-fibre interferometers based on elliptical core fibres \cite{Kim87}, mismatched-core fibres \cite{Can97}, liquid-crystal-clad fibres \cite{Che06}, and photonic crystal fibres \cite{Cho07}.

\section{Two-mode microstructured exposed core fibre}

We consider a standard ECF structure \cite{Sch17} (Fig.~\ref{fig:fig1}) that consists of a Y-shaped silica core (the refractive index $n_{co}=1.4607$, $2.2$~$\mu$m size) formed by three elliptical air holes ($n_{air}=1$) \cite{War09, Kos12, Mes14, Kos14, Sch17}. Two of these holes are fully enclosed by silica, but the third hole is open such that the top surface of the Y-shaped core can be accessed from the outer space across the fibre length. The exposed surface of the Y-shaped hole is covered by a dielectric layer of uniform thickness $h$ and optical refractive index $n_{dl}$. We consider $n_{dl}$ from $1$ to $2.5$ because the dielectric layer deposited on top of a realistic ECF can be made of Teflon ($n_{dl}=1.36$), silica glass ($n_{dl}=1.46$), a polymer ($n_{dl}=1.49$), tellurite glass ($n_{dl}=2$) or other high refractive index materials \cite{Li07, Kos14, Bac19, review}. The optical properties of such fibre originate from the small Y-shaped core with the dielectric layer (inset Fig.~\ref{fig:fig1}) and therefore only this region is considered in our analysis.

\begin{figure}[t]
\centerline{
\includegraphics[width=7.5cm]{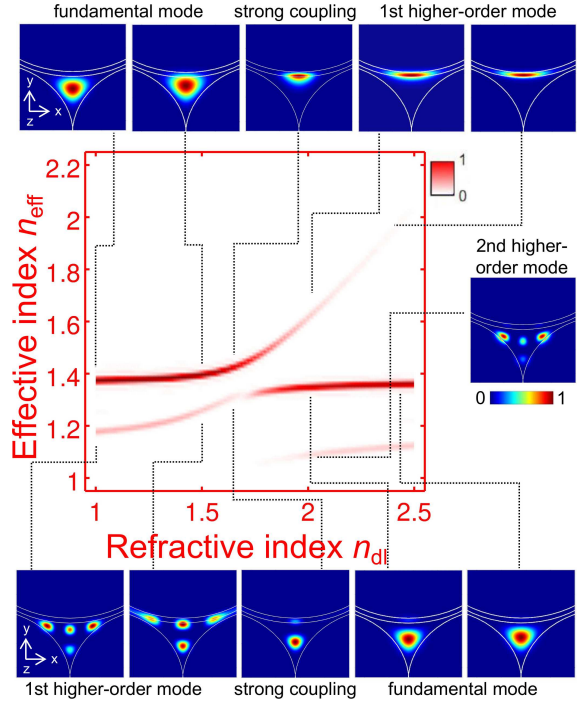}}
\caption{Dispersion characteristic of the ECF with the $100$-nm-thick dielectric layer. The colour map is composed of individual power density spectra calculated for $n_{dl}=1-2.5$. The peaks in these spectra trace dispersion curves, but their magnitude corresponds to the relative fraction of the power in each guided mode (encoded as the intensity of red colour). The insets show the modal intensity profile in the $xy$-plane of the fibre whose contours are outlined by the white curves.}
\label{fig:fig2}
\end{figure}

Figure~\ref{fig:fig2} shows the dispersion characteristic of the ECF and the representative modal optical intensity profiles in the $xy$-plane. An finite-difference mode solver was used \cite{Mak04}. The wavelength is $532$\,nm and the thickness of the dielectric layer is $h=100$\,nm. Qualitatively similar results were obtained for all wavelengths across the visible spectral range as well as for the fibres with the other experimentally accessible \cite{Kos14} dielectric layer thicknesses $50-200$\,nm. 

The bare fibre ($n_{dl}=1$) supports two guided modes with the effective refractive indices $n_{eff}=1.183$ and $n_{eff}=1.377$. The mode with $n_{eff}=1.377$ is the fundamental mode and the other mode is a higher-order guided mode. 

At $n_{dl}=1$, the two modes behave independently -- their resonance peaks are separated and have different magnitudes and linewidths. However, because of a strong dependence of $n_{eff}$ of the higher-order mode on $n_{dl}$, the two modes hybridise at $n_{dl} \approx 1.7$, which is evidenced by hybridisation of their respective modal intensity profiles (Fig.~\ref{fig:fig2}), nearly equal magnitude of their power density peaks, and avoided dispersion curve crossing \cite{Tan18_1}.

At $n_{dl}>1.8$, the hybridisation disappears and the two modes again become independent resonators. The fundamental mode regains a profile similar to that of the bare fibre. However, the profile of the higher-order mode becomes completely different -- the light is localised in the dielectric layer where the refractive index is higher than that of the core.

At $n_{dl} > 1.7$ there also appears the second higher-order mode with the light localised at the interface between the Y-shaped core and air. Initially, the fraction power carried by this mode is low, but it gradually increases as the value of $n_{dl}$ is increased.      

In the following, we will consider the regime of $n_{dl} > 1.8$ where the guided modes strongly confine light in the core and the dielectric layer. We will not consider the regime of mode anti-crossing ($n_{dl} \approx 1.7$), although that regime would be suitable for other applications \cite{Tan18_1}. 

\section{Analytical model of sensitivity}

{\color{black}As the light propagates along the fibre section with the dielectric layer of length $\ell$ [Fig.~\ref{fig:fig1}(c)], a phase difference arises from the difference in propagation constants of the fundamental mode in the Y-shaped core, $\beta_{co}$, and the higher-order mode in the dielectric layer, $\beta_{dl}$. A fraction of the power in the higher-order mode can be coupled back into the fundamental mode. The out-of-phase component results in an interference effect that can be observed as an oscillating attenuation measured as a function of wavelength at the output edge of the fibre.}  

We use the perturbation theory \cite{Snyder} where we assume that the refractive index of the dielectric layer is perturbed as $n_{dl} + \Delta n_{dl}$, but that of the fibre core remains unchanged. This idealised model allows us to validate the analytical formalism presented below and also corresponds to a realistic scenario of temperature sensing \cite{Li07}.  

The phase of fringes arising due to the interference effect is $\phi = \frac{2\pi \ell \Delta n_{eff}}{\lambda} = \ell \Delta \beta$, where $\lambda$ is the wavelength of light in free space, $\ell$ is the length of the fibre section with the dielectric layer [Fig.~\ref{fig:fig1}(c)], and $\Delta \beta {=} \beta_{co} {-} \beta_{dl}$ is the difference in the propagation constants of the interfering modes. Under the condition of sufficiently small $\Delta n_{dl}$ we obtain $\beta_{dl} = \bar{\beta}_{dl} + k \bar{\eta}_{dl} \Delta n_{dl}$, where $\bar{\beta}_{dl}$ and $\bar{\eta}_{dl}$ are the unperturbed propagation constant and the fraction power of the mode guided in the dielectric layer, respectively, and $k$ is the wavevector. Thus, by following \cite{Can97}, the refractive index change $\Delta n_{dl}$ can be related to the measurable change in phase $\Delta \phi$ as $\Delta n_{dl} \approx \frac{\lambda \Delta \phi}{2\pi \bar{\eta}_{dl} \ell}$. 

\begin{figure}[t]
\centerline{
\includegraphics[width=9.5cm]{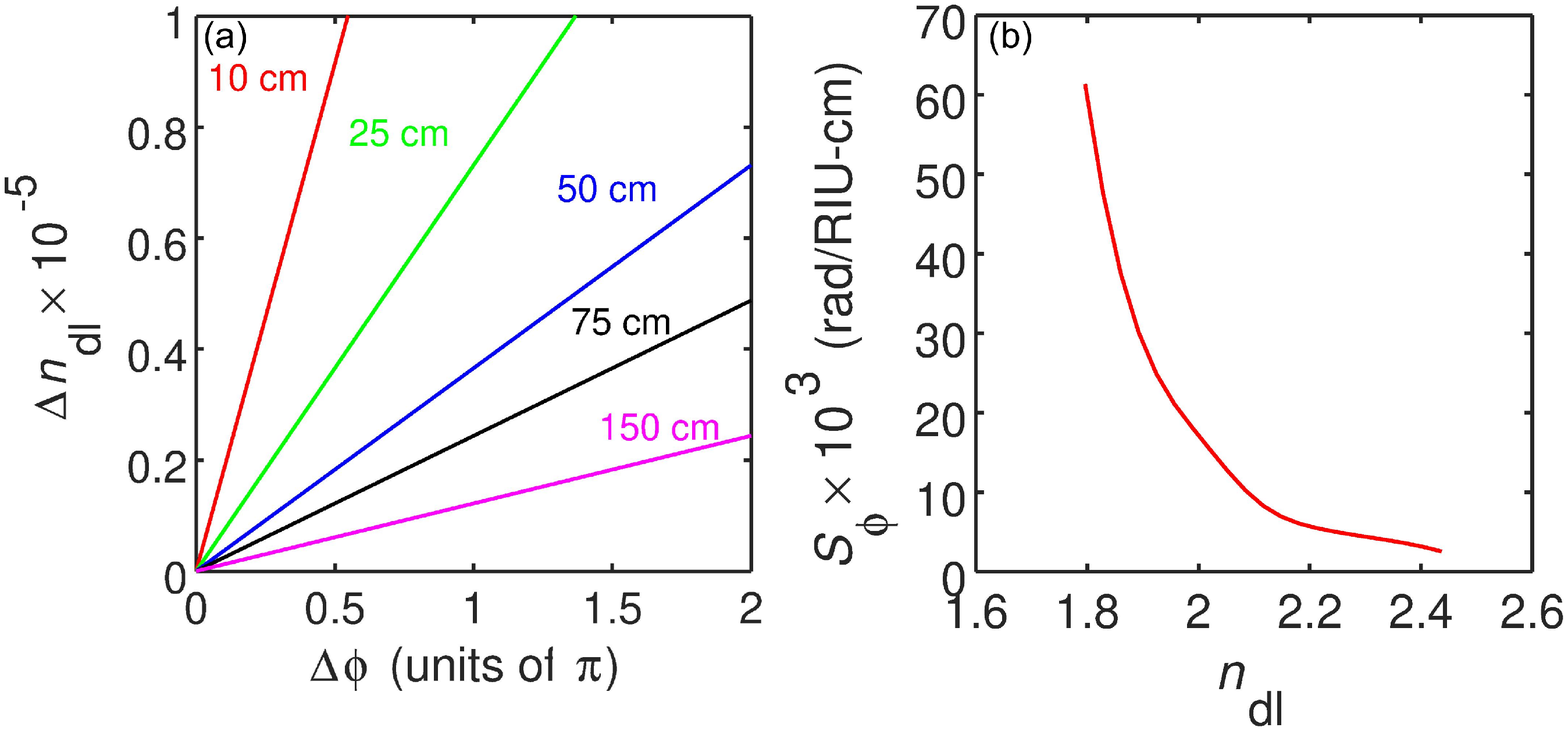}}
\caption{(a) Change in the optical refractive index of the dielectric layer of the ECF, $\Delta n_{dl}$, as a function of the phase change $\Delta \phi$ for several representative lengths $\ell$ {\color{black} defined in Fig.~\ref{fig:fig1}(c)}. The thickness of the dielectric layer is $100$\,nm and $n_{dl} = 2$. For example, at $\ell = 150$\,cm one could sense refractive index changes of ${\sim} 10^{-6}$ leading to the change of phase $\Delta \phi = \pi$. (b) Phase sensitivity as a function of $n_{dl}$.}
\label{fig:fig3}
\end{figure}

In Fig.~\ref{fig:fig3}(a) we plot $\Delta n_{dl}$ as a function of $\Delta \phi$ for several representative fibre section lengths $\ell$. We consider a dielectric layer material with $n_{dl}=2$, which in a practice would correspond to tellurite glass \cite{Li07}. We observe that by using a sufficiently long section of the fibre, one could sense refractive index changes of ${\sim} 10^{-6}$ by using a single interferometer arm for the interfering modes and without disturbing the mode propagating in the core of the fibre.

We also calculate the phase sensitivity \cite{Can97, Qin16} as $S_{\phi} = \frac{\Delta \phi}{\Delta n_{dl} \ell}$. For the practical scenario \cite{Can97, Li07} of $\Delta \phi = \pi$ for $n_{dl} = 2$ we obtain $S_{\phi} \approx 17,000$\,rad/RIU-cm. This value is of the same order of magnitude as $S_{\phi}$ predicted for a slow-light Mach–Zehnder interferometer \cite{Qin16}. In Fig.~\ref{fig:fig3}(b), we plot $S_{\phi}$ as a function of $n_{dl}=1.8...2.4$. The model predicts a gradual decrease in $S_{\phi}$ from $60,000$\,rad/RIU-cm to $3,000$\,rad/RIU-cm, which is consistent with the diverging behaviour of the dispersion curves of the fundamental and the first higher-order modes (Fig.~\ref{fig:fig2}).       

\section{Rigorous numerical simulations}

We numerically verify the analytically predicted sensitivity in Fig.~\ref{fig:fig3}. FDTD simulations of centimetre-long fibre sections are very impractical due to prohibitive computational resource requirements. Therefore, in our analytical model we simultaneously increase $\Delta n_{dl}$ and decrease $\ell$ to the micron range to maintain the same value of $S_{\phi}$. We make sure that $\Delta n_{dl}$ remains small enough not to disrupt the validity of the approximations made in the analytical model. Thus, we choose $\ell=50$\,$\mu$m and $\Delta n_{dl} \approx 3.66 \times 10^{-2}$.    

\begin{figure}[t]
\centerline{
\includegraphics[width=8.5cm]{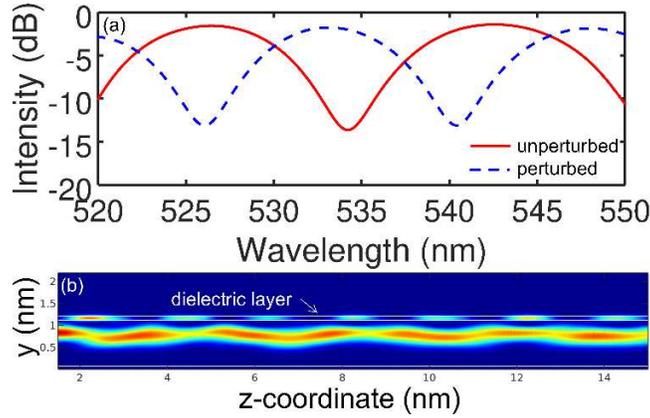}}
\caption{ (a) Calculated fringe pattern produced by the $\ell=50$\,$\mu$m fibre section with the dielectric layer {\color{black}[$\ell$ is defined in Fig.~\ref{fig:fig1}(c)]} with the unperturbed refractive index of the dielectric layer ($n_{dl}=2$, solid curve) and the perturbed refractive index ($n_{dl}+\Delta n_{dl}$, dashed curve). In agreement with the analytical theory, the phase shift $\Delta \phi \approx \pi$ is observed at $\sim 532$\,nm. (b) Optical intensity distribution in a short region of the fibre. }
\label{fig:fig4}
\end{figure}

Figure~\ref{fig:fig4}(a) shows the fringe pattern calculated by collecting and Fourier-transforming the light emitted from the output edge of the fibre [Fig.~\ref{fig:fig1}(c)]. The observed fringe shift $\Delta \phi \approx \pi$ at $\sim 534$\,nm is in good agreement with the predictions of the analytical theory. The small shift of the fringe from the nominal wavelength $532$\,nm is an artefact.

Figure~\ref{fig:fig4}(b) shows the simulated optical intensity profile along the length of the fibre. We observe a picture predicted by the modal simulations in Fig.~\ref{fig:fig2} -- the light is guided through the fibre core and the dielectric layer, and the optical energy is periodically exchanged between the fibre core and the dielectric layer. In the fibre core, the coupling leads to a shift of the fundamental mode profile closer to the dielectric layer, which results in a zigzag pattern extended along the $z$-coordinate. The light localisation in the dielectric layer follows this zigzag behaviour by creating a chain of equidistant hotspots. The spacing between these hotspots equals ${\sim}2.0$\,$\mu$m and it is in good agreement with the predicted period of interference effect at $\lambda=532$\,nm and $n_{dl}=2$ calculated as $\lambda/\Delta n_{eff}$ using the data from Figure~\ref{fig:fig2}.                     

\begin{figure}[t]
\centerline{
\includegraphics[width=8.5cm]{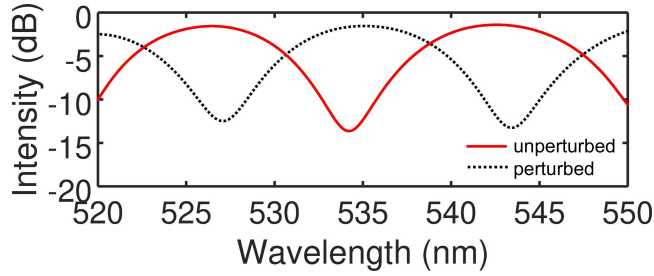}}
\caption{Fringe pattern produced by the $50$-$\mu$m-long fibre section with the dielectric layer with the unperturbed (solid curve) and uniformly perturbed (dotted curve, $\Delta n_{out} \approx 9.2 \times 10^{-2} $) refractive index of the medium located above the dielectric layer. The refractive indices of the core and dielectric layer, $n_{dl}=2$, are constant.}
\label{fig:fig5}
\end{figure}

We also show that the investigated ECF is highly sensitive to changes in the refractive index, $n_{out}$, of the medium located above the dielectric layer. In this case, the refractive indices of both fibre core and dielectric layer remain unchanged, but that of the outer medium is perturbed as $n_{out} + \Delta n_{out}$. This scenario is difficult to analyse analytically because of the need to calculate leaky-mode losses \cite{Cho08}. Hence, we employ  the FDTD method. Figure~\ref{fig:fig5} shows the fringe patter calculated for a $50$-$\mu$m-long fibre section with the dielectric layer. We use $\Delta n_{out} = 9.2 \times 10^{-2}$ and $n_{dl}=2$. We observe a phase shift of $\sim \pi$ and we obtain $S_{\phi} = 6,830$\,rad/RIU-cm. 

\section{Conclusion}
Our analytical modelling and numerical simulations have demonstrated that we can make a single-arm interferometric sensor by coating the core of a microstructured exposed core fibre with a high refractive index material. We show that the proposed fibre structure can guide light in both the core and coating, thereby satisfying conditions for strong mode interference and providing the sensitivity of up to $60,000$\,rad/RIU-cm. Single-arm fibre-optical interferometers are simpler and more robust than those based on two arms, and therefore the investigated fibre structure should be especially suitable for sensing of small refractive index shifts induced by temperature changes, chemical reaction or certain diseases in living organisms. 

\section*{Funding}
\noindent Australian Research Council (ARC) Future Fellowship (FT180100343, FT160100357), Centre of Excellence for Nanoscale BioPhotonics (CE140100003) and LIEF (LE160100051) programs.  




\end{document}